\documentstyle[prl,aps]{revtex}
\begin{document}
\title{\large \bf 
Evolution of Correlation Properties and Appearance of Broken \\
Symmetry in the Process of Bose-Einstein Condensation}

\author {Yu.\ Kagan and B.V.\ Svistunov}
\address{
Russian Research Center ``Kurchatov Institute", 123182 Moscow, Russia}

\maketitle
\begin{abstract}
We consider the self-evolution of strongly non-equilibrium interacting
Bose gas. Due to the mere fact of large (as compared to unity) occupation 
numbers in the initial state the problem is directly reduced to the 
question of temporal evolution of the statistical matrix diagonal in the 
coherent-state representation.

Strictly speaking, gauge symmetry is not destroyed even when the 
long-range ordering is completed, owing to the ineviatble averaging over 
the ensemble. Actual symmetry breaking can occur only as a result of 
introducing a small term of the Hamiltonian violating conservation of 
particles, or quantum-mechanical measurement, also implying 
non-conservation of particles.
\vspace{0.2cm} 
\vspace{0.5cm}
\end{abstract}

Kinetics of the Bose-Einstein condensate formation from a strongly
non-equilibrium (completely disordered) initial state, ultimately leading 
to the long-range ordering, is one of the most interesting and fundamental
dynamic problem of the physics of multi-particle systems.
Recent exciting experiments on Bose-Einstein condensation (BEC)
in ultra cold trapped atomic gases \cite{BEC} open up a unique 
opportunity of experimental study of this problem. Indeed, magnetic
confinement completely isolates the system from the walls
leading to the most refined situation, when the 
kinetics is the result of pure self-evolution of non-equilibrium gas.
On the other hand, (forced) evaporative cooling technique allows to
have a variety of initial conditions differing by the number of 
particles and the residual energy. The formation of the correlation
properties can be traced non-destructively by measuring the secondary
particles in the nonelastic processes, optical or dynamic 
response of the system.

Suppose that upon a fast cooling down there arises an isolated system
with a certain fixed number of particles, $N$, the energy per particle,
$\varepsilon_*$, being considerably less than the BEC temperature, 
$T_c$, that corresponds to the given particle density,
$n$. When such a system eventually comes to equilibrium in the course of
self-evolution, the number of condensate particles will be close to $N$. 
The sharp contrast between this final state and condensateless initial
one predetermines rather rich and nontrivial evolution scenario,
involving a number of qualitatively different stages \cite{Sv,KSS,KS}.

The particular analysis of the kinetics naturally starts with the 
statement that the evolution of the weakly interacting Bose gas with
large (as compared to unity) occupation numbers is essentially 
classical-field phenomenon and thus can be described from the very 
beginning in terms of non-linear Scr\"{o}dinger equation. Taking this 
for granted, in Refs.\ \cite{Sv,KSS,KS} a detailed scenario of the BEC 
process was developed (see also Ref.\ \cite{K}). However, the very fact of
possibility of the pure classical-field description is sometimes 
questioned. The heart of the doubts is the prejudice that for 
many-particle Bose systems the validity of the classical-field 
description implies the presence of the condensate. Recently, a special 
effort has been undertaken by Stoof \cite {St} in order to substantiate 
the viewpoint that prior 
to the onset of the classical-field regime there should take place a sort 
of essentially quantum phase transition - ``nucleation of condensate". 
(Published just now Ref.\ \cite{St2} also contains the
affirmation that there takes place a phase transition at an intermediate 
moment of evolution.)

In this Letter we present a direct proof of the statement that the only 
fact of large (as compared to unity) occupation numbers in the completely 
disordered initial state forms the enough condition for reducing the 
problem to the question of temporal evolution of the statistical matrix 
diagonal in the coherent-state representation. This immediately leads to 
the substantiation of the classical-field description: At large occupation 
numbers the coherent states of the initial ensemble evolve following
``independent trajectories".  Typical members of the ensemble are basically 
similar to each other, so for all practical purposes it is enough to consider 
the evolution of just one generic member of the ensemble. It is 
essential that though the eigenvalue of the field operator is non-zero 
for any particular member of the ensemble (and obeys non-linear 
Schr\"{o}dinger equation), this by no means implies the presence of the
condensate.

Discussing the evolution of the correlation properties,
we point out that the size of ``long-range-ordered" region 
increases gradually with the suppression
of the fluctuations of the phase, until the size of the system 
is reached. Thus we make sure that the condensation kinetics is 
a cross-over process, which does not imply any phase transition at 
some intermediate stage of evolution. 

Though one deals with an ensemble of states which are not 
eigenfunctions 
with respect to the gauge transformation of the
field operator (virtual symmetry breaking), the genuine gauge symmetry
is not destroyed, owing to the averaging over the ensemble.
Symmetry breaking can occur only as a result of
introducing a small term of the Hamiltonian violating conservation
of particles, or quantum-mechanical measurement, also 
implying non-conservation of particles.

Let us turn to the formal analysis. For simplicity, we consider a 
homogeneous case. The parameter $na^3$ ($a$ is the scattering length)
is assumed to be small enough to guarantee that the above-mentioned
condition
\begin{equation}
\varepsilon_* \ll T_c 
\label{cond1}
\end{equation}
holds true along with the condition ($\hbar = 1$)
\begin{equation}
nU_0 \ll \varepsilon_* \; \;  , ~~~~~~ 
U_0=\frac{4 \pi a }{m} \; \; .
\label{cond2}
\end{equation}
This means that at the initial moment most of the particles are in
the kinetic region, where non-equilibrium state can be 
described in terms of distribution of occupation numbers.
Clearly, for the characteristic occupation number, $n_{k_*}$,
($k_*^2/2m = \varepsilon_*$) we have
\begin{equation}
n_{k_*} \sim \left( \frac{3T_c}{\varepsilon_*}\right)^{3/2} \gg 1 \; \; .
\label{cond3}
\end{equation}

To get a preliminary insight into the structure of the initial state,
it is extremely instructive to estimate the density-density correlation
function, $K(r)= \langle \hat{n}(0) \hat{n}({\bf r}) \rangle$, 
where $\hat{n}({\bf r})$ is the number-density operator at the point
${\bf r}$. As can be easily shown with the help of Wick's theorem,
$K(r)$ varies from $2n^2$ (at $r \rightarrow 0$) to $n^2$ at the distances
of the order of $\lambda \sim 1/k_*$. Together with Eq.\ (\ref{cond3})
implying $\lambda \gg n^{-1/3}$ this means that the density experiences
strong macroscopic fluctuations. Thus we realize that from the very
beginning we deal with a sort of turbulent state with strong local 
non-uniformity.

Further proceeding requires an explicit form of the statistical 
matrix. At the beginning, the statistical matrix of the whole 
system, $W$, is equal, to a good approximation, to the direct
product of density matrices, $W_{\bf k}$, for each one-particle momentum 
state. The latter are diagonal in the occupation number representation:
\begin{equation}
W \, = \, \prod_{\bf k} W_{\bf k} \; \; , ~~~
W_{\bf k} \, = \, \sum_{n_{\bf k}} Q_{\bf k}(n_{\bf k}) \,
\mid n_{\bf k} \, \rangle \langle \, n_{\bf k} \mid \; \; .
\label{W}
\end{equation}
Since the coherent states 
\begin{equation}
\mid \alpha_{\bf k} \, \rangle = 
\exp (-\frac{1}{2} \mid \alpha_{\bf k} \mid^2 \, ) \, 
\sum^{\infty}_{n_{\bf k}=0} \frac{(\alpha_{\bf k})^{n_{\bf k}}}
{\sqrt{n_{\bf k} !}} \, \mid n_{\bf k} \, \rangle 
\label{CS}
\end{equation}
form a complete set \cite{Gl}, we can rewrite Eq.\ (\ref{W}) in the 
coherent-state representation. Assuming that $Q_{\bf k}$ is 
a smooth quasi-continuous function of $n_{\bf k}$ (a natural property 
of completely disordered initial state), one can prove that, 
up to smaller corrections in the parameter $ 1/\bar{n}_{\bf k}$,
$W_{\bf k}$ is diagonal in new representation, satisfying the relation
\begin{equation}
W_{\bf k} \, = \, \int d (\rho_{\bf k}^2) \, 
\int \frac{d \varphi_{\bf k}}{2 \pi} \,
Q_{\bf k}( \rho_{\bf k}^2)
\mid \alpha_{\bf k} \, \rangle
\langle \, \alpha_{\bf k} \mid \; \; ,
\label{main1}
\end{equation}
where $\alpha_{\bf k} = \rho_{\bf k} \exp (i \varphi_{\bf k})$.
The most transparent way of proving this important result is to
return from (\ref{main1}) to (\ref{W}). Let us substitute Eq.\ (\ref{CS}) 
into the right-hand side of the Eq.\ (\ref{main1}) and perform the 
integration over $\varphi_{\bf k}$. This leads to the expression 
diagonal in the occupation number representation:
\begin{equation}
\int d (\rho_{\bf k}^2) \, Q_{\bf k}( \rho_{\bf k}^2) \,
\sum_{n_{\bf k}} \frac{1}{n_{\bf k} !} \, 
\rho_{\bf k}^{2n_{\bf k}} \, e^{-\rho_{\bf k}^2} \,
\mid n_{\bf k} \, \rangle
\langle \, n_{\bf k} \mid \; \; .
\label{expr}
\end{equation}
Then change the order of the summation and integration in the expression
(\ref{expr}) and notice, that the smoothness of $Q_{\bf k}$ and large
typical $n_{\bf k}$'s allow replacing
\begin{equation}
\rho_{\bf k}^{2n_{\bf k}} \, e^{-\rho_{\bf k}^2} \, \rightarrow \,
n_{\bf k} ! \, \delta(\rho_{\bf k}^2 - n_{\bf k}) \; \; ,
\label{repl}
\end{equation}
immediately leading to Eq.\ (\ref{W}). 

Now we introduce the wavefunction of the coherent state of the system
(at $t=0$, $\alpha \equiv \{ \alpha_{\bf k} \} $)
\begin{equation}
\Phi (0,\alpha) \, = \, \prod_{\bf k}
\mid \alpha_{\bf k} \, \rangle \; \; .
\label{coh}
\end{equation}
This function is an eigenfunction of the field operator
$\hat{\psi}$ with the eigenvalue
\begin{equation}
\psi \, = \, \sum_{\bf k} \rho_{\bf k} e^{i \varphi_{\bf k}} 
e^{i \bf k r} \; \; .
\label{eig}
\end{equation}
In terms of coherent states (\ref{coh}), the statistical matrix of the
initial state takes the form
\begin{equation}
W(0) \, = \, \int {\cal D} \alpha  \, 
Q ( \alpha ) \,
\mid \Phi (0, \alpha)  \, \rangle
\langle \, \Phi (0, \alpha) \mid \; \; ,
\label{main2}
\end{equation}
where 
\begin{equation}
{\cal D} \alpha \, = \, \prod_{\bf k} d (\rho_{\bf k}^2) \,
\frac{d \varphi_{\bf k}}{\pi} \; \; ,~~~~~~ 
Q(\alpha)=\prod_{\bf k} Q_{\bf k} (\rho_{\bf k}^2) \; \; .
\end{equation}
Thus we see that at the beginning the statistical matrix of the system
is diagonal in the coherent-state representation. Note, that $Q$ is
independent of $\varphi_{\bf k}$'s. This means that the phases of
harmonics in the initial state are completely non-correlated.

Consider the evolution of the wavefunction $\Phi (t)$, which at the 
initial moment coincides with some coherent state $\Phi (0,\alpha)$
After small temporal interval $\Delta t$ we have
\begin{equation}
\Phi (\Delta t) \, = \, e^{-iH \, \Delta t} \Phi (0) \, = \, 
\Phi (0) - i \Delta t H \, \Phi (0)  \; \; .
\label{Phi}
\end{equation}
Here $H$ is the Hamiltonian of the interacting Bose gas (the external
potential $V({\bf r})$ is introduced for generality)
\begin{equation}
H \, = \, \int d{\bf r} \, \hat{\psi}^{\dag} 
[-\frac{\Delta}{2m} + V({\bf r})] \hat{\psi}
+ \frac{U_0}{2} \int d{\bf r} \, 
\hat{\psi}^{\dag} \hat{\psi}^{\dag} \hat{\psi} \, \hat{\psi} \; \; .
\label{Ham}
\end{equation}
Let us act by the field operator on $\Phi (\Delta t)$. Since by definition
$\hat{\psi} \Phi (0) = \psi (0) \Phi (0)$, from Eq.\ (\ref{Phi}) we have
\begin{equation}
\hat{\psi} \Phi (\Delta t) \, = \, \psi (0) \Phi (0) - 
i \Delta t \psi(0) H \Phi (0) -
i \Delta t [\hat{\psi}, H] \Phi (0) \; \; .
\label{rel1}
\end{equation}
With the explicit form of the commutator,
$[\hat{\psi}, H] \, = \, -(1/2m) \Delta \hat{\psi} + V \hat{\psi} + 
U_0 \hat{\psi}^{\dag} \hat{\psi} \, \hat{\psi}$,
the only nontrivial term in Eq.\ (\ref{rel1}) is 
$\hat{\psi}^{\dag} \hat{\psi} \, \hat{\psi} \Phi (0) \, = \, 
\psi^2(0) \hat{\psi}^{\dag} \Phi (0)$.
As can be checked straightforwardly, taking into account the smooth
and rather narrow distribution of $n_{\bf k}$ around 
$\bar{n}_{\bf k}$ in the $\alpha_{\bf k}$-state with
$\rho^2_{\bf k}=\bar{n}_{\bf k} \gg 1$, we have
\begin{equation}
\hat{\psi}^{\dag} \Phi (0) \, = \, 
\psi^*(0) \Phi (0) + {\cal O} (1/ \bar{n}_{\bf k}) \; \; .
\label{rel3}
\end{equation}
So finally we obtain
\begin{equation}
\hat{\psi} \Phi (\Delta t) \, = \, 
 \left[ \psi(0) - i \Delta t \left( 
-\frac{\Delta}{2m} \psi (0) + V \psi (0) 
 + U_0 \mid \psi (0)\mid^2 \psi (0) \right) \right] \Phi (\Delta t) 
+ {\cal O} (1/n_{\bf k}) + {\cal O} ((\Delta t)^2)  \; \; .
\label{rel4}
\end{equation}
Repeating iteratively the above procedure and taking the limit 
$\Delta t \rightarrow 0$ reveals that, apart from the $1/\bar{n}_{\bf k}$ 
corrections,  $\Phi (t)$ is a coherent state,
\begin{equation}
\hat{\psi} \Phi (t) \, = \, \psi (t) \Phi (t) \; \; ,
\label{rel5}
\end{equation}
with eigenvalue $\psi (t)$ obeying non-linear Schr\"{o}dinger equation (NSE)
\begin{equation}
i \frac{\partial \psi}{\partial t} \, = \, 
-\frac{\Delta}{2m} \psi + V \psi + U_0 \mid \psi \mid^2 \psi \; \; .
\label{NSE}
\end{equation}

Hence, each separate coherent state of the ensemble evolves following
an ``independent trajectory" (cf. Ref.\ \cite{Carr}), and
for the statistical matrix we thus have
\begin{equation}
W(t) \, = \, \int {\cal D} \alpha  \, 
Q ( \alpha ) \,
\mid \Phi (t, \alpha)  \, \rangle
\langle \, \Phi (t, \alpha) \mid \; \; .
\label{SO}
\end{equation}
Diagonal statistical matrix (\ref{SO}) with $Q(\alpha) > 0$ describes
the temporal evolution of a statistical ensemble of coherent states
$\Phi(t,\alpha)$. Dealing with essentially macroscopic system, almost
for all practical purposes it is enough to consider the evolution of 
some typical representative of the ensemble with a particular choice
of $\alpha \equiv \{ \varphi_{\bf k}, \rho_{\bf k} \}$. The solution
of Eq.\ (\ref{NSE}) for the initial state (\ref{eig}) defined by this 
set of parameters gives us the classical-field picture of self-evolution.
It is worth noting, that calculating any correlation function with 
$W(t)$ (\ref{SO}), one first performs averaging over a coherent state 
and only then averaging over the ensemble. Eventually, most of the
gross-features prove to be insensitive to the initial parameters 
just after the first averaging.

The average of the operator $\hat{\psi}$ over the coherent state,
\begin{equation}
\langle \, \hat{\psi} \, \rangle \, = \, 
\langle \, \Phi(t,\alpha) \mid \hat{\psi} \mid \Phi(t,\alpha) \, \rangle \,
= \, \psi(t, \alpha) \; \; ,
\label{av1}
\end{equation}
is non-zero and is defined by the solution of NSE (\ref{NSE}).
However, ensemble averaging with $W(t)$ (\ref{SO}) leads to
\begin{equation}
\langle \! \langle\, \hat{\psi} \, \rangle \! \rangle \, \equiv \, 
\psi_0 \, = \, 0  \; \; .
\label{av2}
\end{equation}
Eq.\ (\ref{NSE}) and virtual broken symmetry for any component of the 
ensemble do not imply the presence of the condensate. Eq.\ (\ref{av2})
reflects global gauge symmetry preserving during all the evolution.

Consider one-particle density matrix 
\begin{equation}
K({\bf r},t) \, = \, \frac{1}{\Omega} \int d {\bf r}' \,
\langle \! \langle \, \hat{\psi}^{\dag} ({\bf r}+{\bf r}') \, 
\hat{\psi} ({\bf r}')
 \, \rangle \! \rangle  \; \; .
\label{DM}
\end{equation}
($\Omega$ is the volume of the system.)
Averaging over a coherent state is straightforward:
\begin{equation}
K_{\alpha}({\bf r},t) \, = \, \frac{1}{\Omega} \int d {\bf r}' \,
\langle \, \Phi(t,\alpha ) \mid
\hat{\psi}^{\dag} ({\bf r}+{\bf r}') \, \hat{\psi} ({\bf r}')
\mid \Phi(t,\alpha ) \, \rangle   \, = \,
\frac{1}{\Omega} \int d {\bf r}' \, 
\psi^* ({\bf r}+{\bf r}',t,\alpha ) \, \psi ({\bf r}',t,\alpha ) \; \; .
\label{DM1}
\end{equation}
To obtain this correlator, one should find the solution of NSE in
the entire temporal interval. The analysis presented in 
Refs.\ \cite{Sv,KSS,KS} has revealed that the evolution involves 
a number of stages. The process
starts with an explosion-like wave in momentum space propagating
towards the low energy region. At this first stage the potential 
energy is much less than kinetic one, and, taking advantage of the
random-phase approximation, one can go over from the NSE to the
kinetic equation, and to carry out rather detailed analysis, including 
numerical simulations \cite{Sv,ST}. The next stage takes place when
most of the particles, which eventually will form the condensate,
find themselves in the coherent energetic region $\varepsilon < 
\varepsilon_c = n_0 U_0$. In this region the potential energy becomes
greater than the kinetic one, and NSE thus cannot be reduced to
the Boltzmann equation. Apart from a short transitory period,
the typical momentum of the coherent interval, $k_0(t)$,
is much less than $k_c=\sqrt{mn_0U_0}$. As is shown in 
Ref.\ \cite{KSS}, the solution of Eq.\ (\ref{NSE}), parameterized as
\begin{equation}
\psi \, = \, \sqrt{n} \, e^{i \chi} \; \; ,
\label{psi1}
\end{equation}
is characterized by small density fluctuations, 
$\langle \, (\delta n / n_0)^2 \, \rangle \, \ll \, 1$.
Let us substitute (\ref{psi1}) into (\ref{DM1}):
\begin{equation}
K_{\alpha}({\bf r},t) \, \approx \, \frac{n_0}{\Omega} \int d {\bf r}' \,
e^{ -i \chi ({\bf r}+{\bf r}',t, \alpha) + 
i \chi ({\bf r}',t, \alpha) } \; \; .
\label{K1}
\end{equation}
If at the moment $t$ the characteristic width of the momentum interval
is $k_0(t)$, then at $r \ll r_0(t) = 1/k_0(t)$ the phase difference in
(\ref{K1}) practically vanishes, yielding $K_{\alpha} \approx n_0$. This
result is insensitive to the ensemble averaging (\ref{DM}). Hence, 
at $r < r_0(t)$ the structure of density matrix corresponds to that
of the condensate state. This so-called quasicondensate state does
not, however, posses genuine long-range order: at $r \gg r_0(t)$
the correlator (\ref{K1}) approaches zero. This follows naturally
from the absence of phase correlations at such distances, which
implies automatically anomalous character of long-wave correlations
and the presence of irregular vortex structure \cite{KS}.
The absence of phase correlations means that the correlator
\begin{equation}
\langle \, \nabla \chi ({\bf r}) \, \nabla \chi (0) \, \rangle \, 
\approx \, \sum_{k<k_0(t)} k^2 \mid \chi_{\bf k} \mid^2 e^{i \bf kr} \; \; .
\label{chixchi}
\end{equation}
vanishes at $r \gg r_0(t)$. Since the decay of correlations takes place
at the scale $\sim r_0(t)$, and is associated with the typical phase
difference $\sim \pi$, one estimates ($\gamma \sim 1$)
\begin{equation}
\mid \chi_{\bf k} \mid^2 \, \approx \, \frac{\gamma}{\Omega k^3} \; \; ,
~~~~ k < k_0(t) \; \; .
\label{chi2}
\end{equation}
Expanding phase difference in the exponent of Eq.\ (\ref{K1}) into
the Fourier series, after standard calculations, we have
\begin{equation}
K_{\alpha} = n_0 e^{-S} \; \; , ~~~~ S \, =\,  
\sum_{k<k_0(t)} k^2 \mid \chi_{\bf k} \mid^2 
(1 - \cos {\bf kr}) \; \; .
\label{K3}
\end{equation}
According to (\ref{chi2}), $S$ can be estimated as
\begin{equation}
S \, \approx \, 4\pi \gamma \ln \left( k_0(t) \, r \right) \; \; ,
~~~~  k_0(t) \, r \gg 1 \; \; .
\label{S}
\end{equation}
As fluctuations attenuate, transferring the excess of energy to the
subsystem of normal excitations, $k_0$ monotonically decreases. When
$k_0(t)$ reaches the minimal value $k_{\mbox{\scriptsize min}} \approx 
\pi / L$ ($L$ is the system size), variation of $S$ with $r$ stops
and the correlator $K_{\alpha}$ takes on an independent of $r$ constant
value $n_0$. [Naturally, this result is not affected by averaging with the
statistical matrix (\ref{SO}).] Now the long-range ordering is completed: 
we have the genuine condensate. 

As is discussed in Ref.\ \cite{KS}, the time necessary for this final
stage of evolution depends on what particular process turns out to be 
the ``bottle-neck" for the relaxation kinetics in the case under 
consideration. If this process is the relaxation of the vortex structure,
then the time is $\tau_V \propto L^2$. If the crucial process is the 
relaxation of non-equilibrium long-wave fluctuations of the phase,
then the corresponding time is $\tau_{\varphi} \propto L^2$ 
(hydrodynamic regime), or $\tau_{\varphi} \propto L$ (Knudsen regime)
(see Ref.\ \cite{KS} for the details).

Thus, we have demonstrated that the appearance of the condensate
in the course of self-evolution of non-equilibrium system is an 
essentially cross-over process. The result (\ref{av2}) is preserved
during all the time of evolution, indicating that the global phase of 
the condensate wavefunction is fundamentally unknown. The genuine broken
gauge symmetry (and hence genuine complex order parameter) can appear
at the final step of evolution {\it only} if the Hamiltonian contains
a small term violating conservation of particles, or as a result of
quantum measurement (which also implies exchange of particles in this or
that form).

This work was supported by INTAS (Grant No. INTAS-93-2834-ext),
by NWO (Project No. NWO-047-003.036), and, partially, by the Russian 
Foundation for Basic Research (Grant No. 95-02-06191a).

\end{document}